\documentstyle[twoside,epsf]{article}

\catcode`\@=11
\long\def\@makefntext#1{
\protect\noindent \hbox to 3.2pt {\hskip-.9pt  
$^{{\eightrm\@thefnmark}}$\hfil}#1\hfill}		

\def\@makefnmark{\hbox to 0pt{$^{\@thefnmark}$\hss}}	
	
\def\ps@myheadings{\let\@mkboth\@gobbletwo
\def\@oddhead{\hbox{}
\rightmark\hfil\eightrm\thepage}   
\def\@oddfoot{}\def\@evenhead{\eightrm\thepage\hfil
\leftmark\hbox{}}\def\@evenfoot{}
\def\sectionmark##1{}\def\subsectionmark##1{}}

\oddsidemargin=\evensidemargin
\newcounter{sectionc}\newcounter{subsectionc}\newcounter{subsubsectionc}
\renewcommand{\section}[1] {\vspace{12pt}\addtocounter{sectionc}{1} 
\setcounter{subsectionc}{0}\setcounter{subsubsectionc}{0}\noindent 
	{\tenbf\thesectionc. #1}\par\vspace{5pt}}
\renewcommand{\subsection}[1] {\vspace{12pt}\addtocounter{subsectionc}{1} 
	\setcounter{subsubsectionc}{0}\noindent 
	{\bf\thesectionc.\thesubsectionc. {\kern1pt \bfit #1}}\par\vspace{5pt}}
\renewcommand{\subsubsection}[1] {\vspace{12pt}\addtocounter{subsubsectionc}{1}
	\noindent{\tenrm\thesectionc.\thesubsectionc.\thesubsubsectionc.
	{\kern1pt \tenit #1}}\par\vspace{5pt}}
\newcommand{\nonumsection}[1] {\vspace{12pt}\noindent{\tenbf #1}
	\par\vspace{5pt}}

\newcounter{appendixc}
\newcounter{subappendixc}[appendixc]
\newcounter{subsubappendixc}[subappendixc]
\renewcommand{\thesubappendixc}{\Alph{appendixc}.\arabic{subappendixc}}
\renewcommand{\thesubsubappendixc}
	{\Alph{appendixc}.\arabic{subappendixc}.\arabic{subsubappendixc}}

\renewcommand{\appendix}[1] {\vspace{12pt}
        \refstepcounter{appendixc}
        \setcounter{figure}{0}
        \setcounter{table}{0}
        \setcounter{lemma}{0}
        \setcounter{theorem}{0}
        \setcounter{corollary}{0}
        \setcounter{definition}{0}
        \setcounter{equation}{0}
        \renewcommand{\thefigure}{\Alph{appendixc}.\arabic{figure}}
        \renewcommand{\thetable}{\Alph{appendixc}.\arabic{table}}
        \renewcommand{\theappendixc}{\Alph{appendixc}}
        \renewcommand{\thelemma}{\Alph{appendixc}.\arabic{lemma}}
        \renewcommand{\thetheorem}{\Alph{appendixc}.\arabic{theorem}}
        \renewcommand{\thedefinition}{\Alph{appendixc}.\arabic{definition}}
        \renewcommand{\thecorollary}{\Alph{appendixc}.\arabic{corollary}}
        \renewcommand{\theequation}{\Alph{appendixc}.\arabic{equation}}
        \noindent{\tenbf Appendix \theappendixc #1}\par\vspace{5pt}}
\newcommand{\subappendix}[1] {\vspace{12pt}
        \refstepcounter{subappendixc}
        \noindent{\bf Appendix \thesubappendixc. {\kern1pt \bfit #1}}
	\par\vspace{5pt}}
\newcommand{\subsubappendix}[1] {\vspace{12pt}
        \refstepcounter{subsubappendixc}
        \noindent{\rm Appendix \thesubsubappendixc. {\kern1pt \tenit #1}}
	\par\vspace{5pt}}

\newcommand{\textlineskip}{\baselineskip=13pt}
\newcommand{\smalllineskip}{\baselineskip=10pt}

\def\eightcirc{
\begin{picture}(0,0)
\put(4.4,1.8){\circle{6.5}}
\end{picture}}
\def\eightcopyright{\eightcirc\kern2.7pt\hbox{\eightrm c}} 

\newcommand{\copyrightheading}[1]
	{\vspace*{-2.5cm}\smalllineskip{\flushleft
	{\footnotesize International Journal of Modern Physics B, #1}\\
	{\footnotesize $\eightcopyright$\, World Scientific Publishing
	 Company}\\
	 }}

\newcommand{\publisher}[2]{{\begin{center}\footnotesize\smalllineskip 
	Received #1\\
	Revised #2
	\end{center}
	}}

\def\abstracts#1#2#3{{
	\centering{\begin{minipage}{4.5in}\baselineskip=10pt\footnotesize
	\parindent=0pt #1\par 
	\parindent=15pt #2\par
	\parindent=15pt #3
	\end{minipage}}\par}} 

\renewenvironment{thebibliography}[1]			
	{\frenchspacing
	 \ninerm\baselineskip=11pt
	 \begin{list}{\arabic{enumi}.}
	{\usecounter{enumi}\setlength{\parsep}{0pt}
	 \setlength{\leftmargin 12.7pt}{\rightmargin 0pt} 
	 \setlength{\itemsep}{0pt} \settowidth
	{\labelwidth}{#1.}\sloppy}}{\end{list}}

\newcounter{itemlistc}
\newcounter{romanlistc}
\newcounter{alphlistc}
\newcounter{arabiclistc}

\newcommand{\fcaption}[1]{
        \refstepcounter{figure}
        \setbox\@tempboxa = \hbox{\footnotesize Fig.~\thefigure. #1}
        \ifdim \wd\@tempboxa > 5in
           {\begin{center}
        \parbox{5in}{\footnotesize\smalllineskip Fig.~\thefigure. #1}
            \end{center}}
        \else
             {\begin{center}
             {\footnotesize Fig.~\thefigure. #1}
              \end{center}}
        \fi}

\newcommand{\tcaption}[1]{
        \refstepcounter{table}
        \setbox\@tempboxa = \hbox{\footnotesize Table~\thetable. #1}
        \ifdim \wd\@tempboxa > 5in
           {\begin{center}
        \parbox{5in}{\footnotesize\smalllineskip Table~\thetable. #1}
            \end{center}}
        \else
             {\begin{center}
             {\footnotesize Table~\thetable. #1}
              \end{center}}
        \fi}

\def\@citex[#1]#2{\if@filesw\immediate\write\@auxout
	{\string\citation{#2}}\fi
\def\@citea{}\@cite{\@for\@citeb:=#2\do
	{\@citea\def\@citea{,}\@ifundefined
	{b@\@citeb}{{\bf ?}\@warning
	{Citation `\@citeb' on page \thepage \space undefined}}
	{\csname b@\@citeb\endcsname}}}{#1}}

\newif\if@cghi
\def\cite{\@cghitrue\@ifnextchar [{\@tempswatrue
	\@citex}{\@tempswafalse\@citex[]}}
\def\citelow{\@cghifalse\@ifnextchar [{\@tempswatrue
	\@citex}{\@tempswafalse\@citex[]}}
\def\@cite#1#2{{$\null^{#1}$\if@tempswa\typeout
	{IJCGA warning: optional citation argument 
	ignored: `#2'} \fi}}

\def\pmb#1{\setbox0=\hbox{#1}
	\kern-.025em\copy0\kern-\wd0
	\kern.05em\copy0\kern-\wd0
	\kern-.025em\raise.0433em\box0}

\def\fnt#1#2{\footnotetext{\kern-.3em
	{$^{\mbox{\scriptsize #1}}$}{#2}}}

\def\fpage#1{\begingroup
\voffset=.3in
\thispagestyle{empty}\begin{table}[b]\centerline{\footnotesize #1}
	\end{table}\endgroup}

\def\runninghead#1#2{\pagestyle{myheadings}
\markboth{{\protect\footnotesize\it{\quad #1}}\hfill}
{\hfill{\protect\footnotesize\it{#2\quad}}}}
\font\tenrm=cmr10
\font\tenit=cmti10 
\font\tenbf=cmbx10
\font\bfit=cmbxti10 at 10pt
\font\ninerm=cmr9
\font\nineit=cmti9
\font\ninebf=cmbx9
\font\eightrm=cmr8

\def\qed{\hbox{${\vcenter{\vbox{			
   \hrule height 0.4pt\hbox{\vrule width 0.4pt height 6pt
   \kern5pt\vrule width 0.4pt}\hrule height 0.4pt}}}$}}

\def\bsc{{\sc a\kern-6.4pt\sc a\kern-6.4pt\sc a}}	
\def\bflatex{\bf L\kern-.30em\raise.3ex\hbox{\bsc}\kern-.14em 
T\kern-.1667em\lower.7ex\hbox{E}\kern-.125em X} 

\begin{document}

\runninghead{Tunneling Spectra and Superconducting Gap
in Bi$_{2}$Sr$_{2}$CaCu$_{2}$O$_{8+\delta}$ and 
Tl$_{2}$Ba$_{2}$CuO$_{6+\delta}$} {Tunneling Spectra and 
Superconducting Gap
in Bi$_{2}$Sr$_{2}$CaCu$_{2}$O$_{8+\delta}$ and 
Tl$_{2}$Ba$_{2}$CuO$_{6+\delta}$}

\normalsize\textlineskip
\thispagestyle{empty}
\setcounter{page}{1}

\copyrightheading{}			

\vspace*{0.88truein}

\fpage{1}
\centerline{\bf TUNNELING SPECTRA AND SUPERCONDUCTING GAP }
\vspace*{0.035truein}
\centerline{\bf IN Bi$_{2}$Sr$_{2}$CaCu$_{2}$O$_{8+\delta}$ AND 
Tl$_{2}$Ba$_{2}$CuO$_{6+\delta}$} \vspace*{0.37truein}
\centerline{\footnotesize L.  OZYUZER and J.  F.  ZASADZINSKI}
\vspace*{0.015truein}
\centerline{\footnotesize\it Argonne National Laboratory, Argonne, 
Illinois 60439, USA} 
\baselineskip=10pt 
\centerline{\footnotesize\it 
Illinois Institute of Technology, Chicago, Illinois 60616, USA}
\vspace*{10pt}
\centerline{\footnotesize N.  MIYAKAWA}
\vspace*{0.015truein}
\centerline{\footnotesize\it Science University of Tokyo, Tokyo, Japan}
 
\vspace*{10pt} 
\publisher{31 May 1999}

\vspace*{0.21truein}
\abstracts{Tunneling spectra are reported for 
Bi$_{2}$Sr$_{2}$CaCu$_{2}$O$_{8+\delta}$ (Bi-2212) over a wide doping range 
using 
superconductor-insulator-superconductor (SIS) break junctions.  The 
energy gap inferred from the tunneling data displays a remarkable 
monotonic dependence on doping, increasing to very large values in the 
underdoped region even as T$_{c}$ decreases.  This leads to 
unphysically large values of the strong coupling ratio ($\sim$20).  
The tunneling spectra are qualitatively similar over the entire doping 
range even though the gap parameter, $\Delta$, changes from 12 meV to 60 meV.  
Each spectrum exhibits dip and hump features at high bias with 
characteristic energies that scale with the superconducting gap.  
Tunneling spectra of near optimally-doped 
Tl$_{2}$Ba$_{2}$CuO$_{6+\delta}$ (Tl-2201) also display a weak dip 
feature in superconductor-insulator-normal metal 
(SIN) junctions.  Generated SIS spectra of Tl-2201 are compared with 
measured spectra on Bi-2212 and it is concluded that the dip and hump features 
are generic to high temperature superconductors.  }{}{}

\vspace*{10pt}

\vspace*{12pt}			

\vspace*{1pt}
\textlineskip	
\noindent
High temperature superconductors (HTSs) exhibit a complex 
temperature versus hole doping phase diagram which contains novel 
physics.\cite{emery} Furthermore, the phase diagram is generic for all HTSs 
and is very similar to that found for organic 
superconductors.\cite{organic} This might be an indication that 
electronic features, such as the pseudogap in the underdoped phase of 
cuprates, are universal to layered, correlated systems.  In this paper 
we present measurements of the tunneling density of states (DOS) and 
are particularly focused on high energy features, dip and hump, in the 
single particle excitation spectrum.  
Bi$_{2}$Sr$_{2}$CaCu$_{2}$O$_{8+\delta}$ (Bi-2212) is the most studied 
HTS by surface sensitive probes, because of the clean surfaces which 
arise after cleaving and the availability of crystals over a wide 
doping range.  Tunneling spectroscopy is one of the direct methods 
that measures the DOS and energy gap in superconductors.  
Superconductor-Insulator-Normal metal (SIN) tunneling measurements on 
optimally-doped Bi-2212 have revealed sharp quasiparticle peaks at 
$\mid$eV$\mid$$\sim$$\Delta$, followed by a dip around 2$\Delta$ and a 
broad hump at higher energies most clearly seen on the occupied side 
of the DOS.\cite{yannick,renner} Tunneling data on single crystals of 
Tl$_{2}$Ba$_{2}$CuO$_{6+\delta}$ (Tl-2201) have also exhibited sharp 
quasiparticle peaks, but a relatively weak dip feature that is 
sometimes difficult to see at all in the SIN spectra.\cite{Tl-rapid} 
The precise determination of the energy gap can be more directly 
obtained from Superconductor-Insulator-Superconductor (SIS) junctions.  
Since the resulting tunneling spectrum is the convolution of two 
superconducting DOS, the quasiparticle peaks exactly occur at 
$\pm$2$\Delta$.  Furthermore any fine structure in SIN tunneling 
spectra is enhanced in the SIS geometry along with a shift in energy 
by the gap value.

Tunneling measurements are conducted using {\it Au} tip in a mechanical, 
electrical and magnetic noise isolated environment.\cite{cryo} In this 
study, SIS break junctions of Bi-2212 are obtained by a novel 
technique that is described elsewhere.\cite{yannick} SIS break 
junctions in single crystals of Bi-2212 displayed both quasiparticle 
and Josephson current simultaneously.  However, the techique is not 
successful for Tl-2201 because of strong bonding between planes.  
Neverthless, SIS tunneling spectra of Tl-2201 can be generated from a 
convolution of the SIN tunneling data for a comparison to Bi-2212.


Figure 1(a) shows the doping dependence of Bi-2212 SIS break junction tunneling 
spectra at 4.2 K.  In the figure, each spectrum corresponds to 
different crystals with different hole concentrations, p, from heavily 
overdoped T$_{c}$=56 K, optimally-doped T$_{c}$=95 K, to underdoped 
T$_{c}$=70 K.  Each spectrum is normalized by a constant, shifted 
vertically and Josephson current peak at zero bias deleted for 
clarity.  We note first that the energy gap increases with decreasing doping, 
even as T$_{c}$ drops from the optimally-doped value 95 K down to 70 
K underdoped.  There is thus a clear indication that $\Delta$ does not follow T$_{c}$.  
This raises a question of whether the measured gap is fully due to 
superconductivity or has a contribution from some other effects such 
as spin density wave or charge density wave.\cite{G.D.} Josephson 
tunneling addresses this issue, because multiplication of the 
Josephson current, I$_{c}$, and junction resistance, R$_{n}$, is 
expected to be proportional to the superconducting gap.  The relation between 
I$_{c}$R$_{n}$ and $\Delta$ has shown that the measured gap is 
predominantly due to superconductivity.\cite{miya} This is also 
indicated by the consistent shape of the spectra of Fig.  1(a) over 
the entire doping range.  For all doping levels, the dip and hump 
features are well pronounced and scale with the energy gap.  The dip 
energy, $\omega_{dip}$, is $\sim$3$\Delta$ for optimally-doped 
(consequently 2$\Delta$ in the DOS) and underdoped Bi-2212, and 
approaches 4$\Delta$ for heavily overdoped crystals in SIS 
junctions.  There is also a general weakening of the dip strength as 
doping increases.  Angle resolved photoemission spectroscopy also 
probes single particle excitations, and exhibits spectra around 
($\pi$,0) direction\cite{ding} with similar anomalous dip and hump 
that is found in tunneling.

\begin{figure}[t]
\epsfxsize=12cm 
\centerline{\epsffile{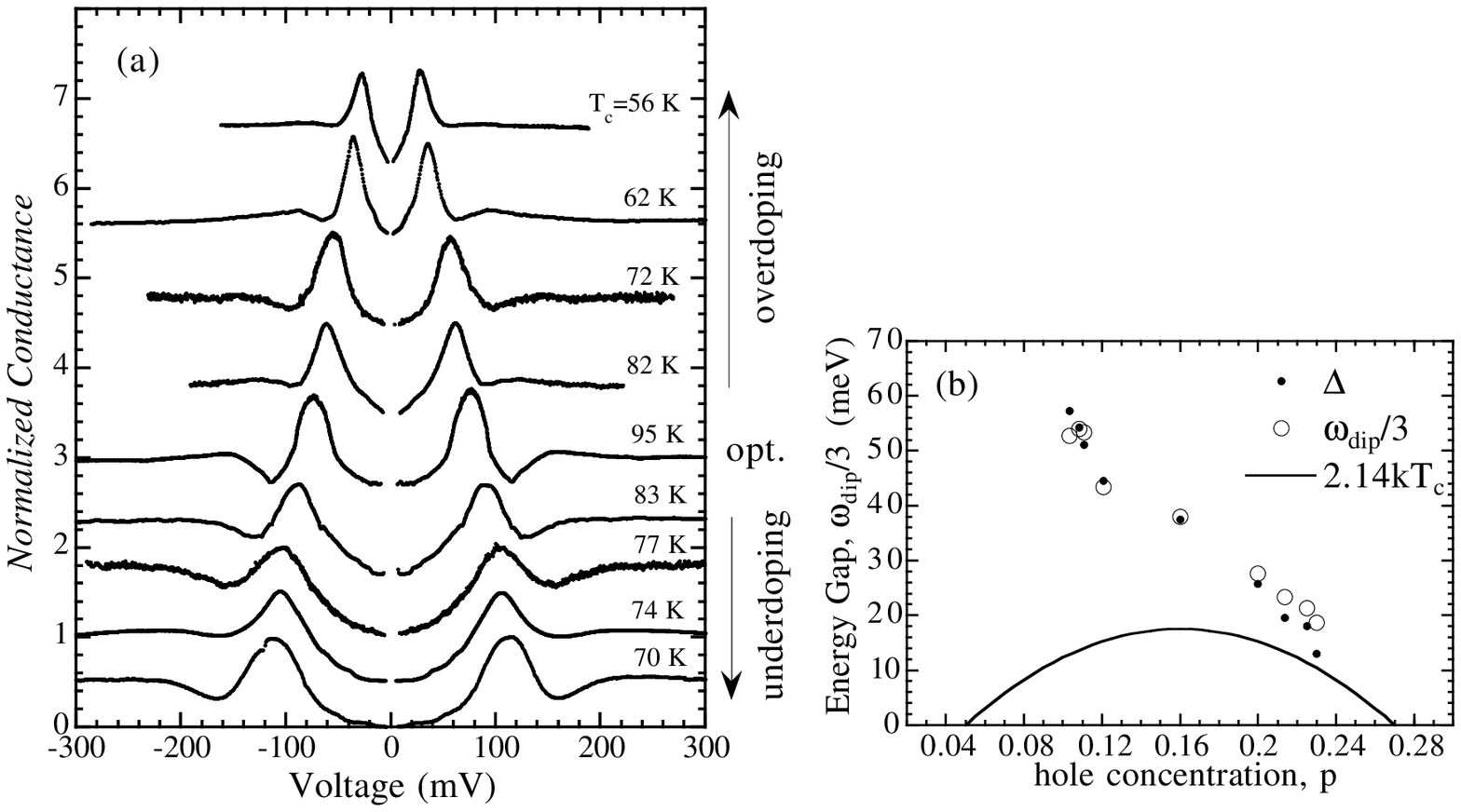}}
\vspace{-9.5cm}
\fcaption{(a) Doping dependence of SIS break junction 
tunneling spectra in Bi-2212.  (b) Energy gap and $\omega_{dip}$/3 versus hole 
concentration from SIS break junctions.} \vspace{-0.3cm}
\end{figure}

Figure 1(b) shows how the measured gap parameter varies with doping in Bi-2212.  
The solid curve is the mean-field gap prediction for $d_{x^{2}-y^{2}}$ 
superconductors, 2.14kT$_{c}$.  The dots are average energy gaps 
obtained on many different junctions and the hole concentrations  are estimated from 
T$_{c}$/T$_{c,max}$=1-82.6(p-0.16)$^{2}$.   
The most underdoped sample with T$_{c}$=70 K exhibits an energy gap 
magnitude of nearly 60 meV which leads to a  strong-coupling 
ratio, 2$\Delta/kT_{c}$$\sim$20.  For the other extreme, a heavily 
overdoped sample with T$_{c}$=56 K, $\Delta$ is around 12-15 meV, and 
this leads to 2$\Delta/kT_{c}$$\sim$5.  Note that the magnitude of energy gap 
approaches the mean field prediction only at the heavily overdoped 
compounds.  To demonstrate the scaling of the dip feature with the 
energy gap, we also plot $\omega_{dip}$/3, for the SIS junctions.  As 
is seen, $\omega_{dip}$ strongly depends on p and scales with 
$\Delta$, furthermore indicating that the phenomenon responsible for 
the dip in heavily overdoped phase is the same as in the underdoped 
phase, and is tied to the superconductivity.

The universal phase diagram of HTSs suggests that the robust dip feature 
in SIS tunneling spectra of Bi-2212 might be seen in other cuprates as 
well.  Tl-2201 junctions have exhibited the most reproducible tunneling data 
that are consistent with a $d_{x^{2}-y^{2}}$ order 
parameter.\cite{Tl-rapid} Since the energy gap magnitude of overdoped 
Bi-2212 with T$_{c}$=62 K is compatible with optimally-doped Tl-2212, 
we compare them in Fig.  2.  We show a set of overdoped Bi-2212 break 
junctions in Fig.  2(a) which exhibit energy gaps between 15-20 meV.  
The SIS spectra also display relativily weak dip and hump feature 
respect to optimally-doped and underdoped Bi-2212 and this is another 
basis for comparison with Tl-2201.  The novel technique mentioned 
earlier has failed to provide SIS break junctions of Tl-2201, however 
SIN tunneling data have been used to generate SIS data which is shown in 
Fig.  2(b).  The convolution of SIN data in Tl-2201 produces the dip 
and hump features that are reasonably consistent with those found on 
overdoped Bi-2212.  However, the locations of the dip features are 
different.  In Tl-2201 the dips are closer to 3$\Delta$ whereas for 
the most overdoped Bi-2212, as mentioned earlier, the dips are closer 
to 4$\Delta$.  This suggests that the strength of the dip may be tied 
to the magnitude of the gap, but that the location is more closely 
tied to the hole concentration.

SIS tunneling spectra of Bi-2212 over a wide doping range and 
generated SIS spectra of optimally-doped Tl-2201 display qualitatively 
similar features, such as sharp quasiparticle peaks, dip and hump 
structures.  This suggests that these higher energy spectral features 
are intrinsic to the DOS of HTS.   In Bi-2212 the dip scales as approximately 
3 times the gap parameter over most of the doping range including the 
underdoped region.  Therefore the difference in energy between the 
dip and the gap also increases as the doping decreases and this 
argues against models of the dip that are associated with the 
collective mode observed in neutron scattering.  The doping dependence of 
the superconducting gap follows that of the pseudogap 
temperature\cite{oda} which suggests that the pseudogap is due to some 
type of precursor superconductivity.
 
\nonumsection{Acknowledgements}
\noindent
This work was partially supported by U.S.  Department of 
Energy, Division of Basic Energy Sciences-Material Sciences under 
contract No.  W-31-109-ENG-38, and the National Science Foundation, 
Office of Science and Technology Centers under contract No.  DMR 
91-20000.

\nonumsection{References}

\begin{figure}[t]
\epsfxsize=13cm 
\centerline{\epsffile{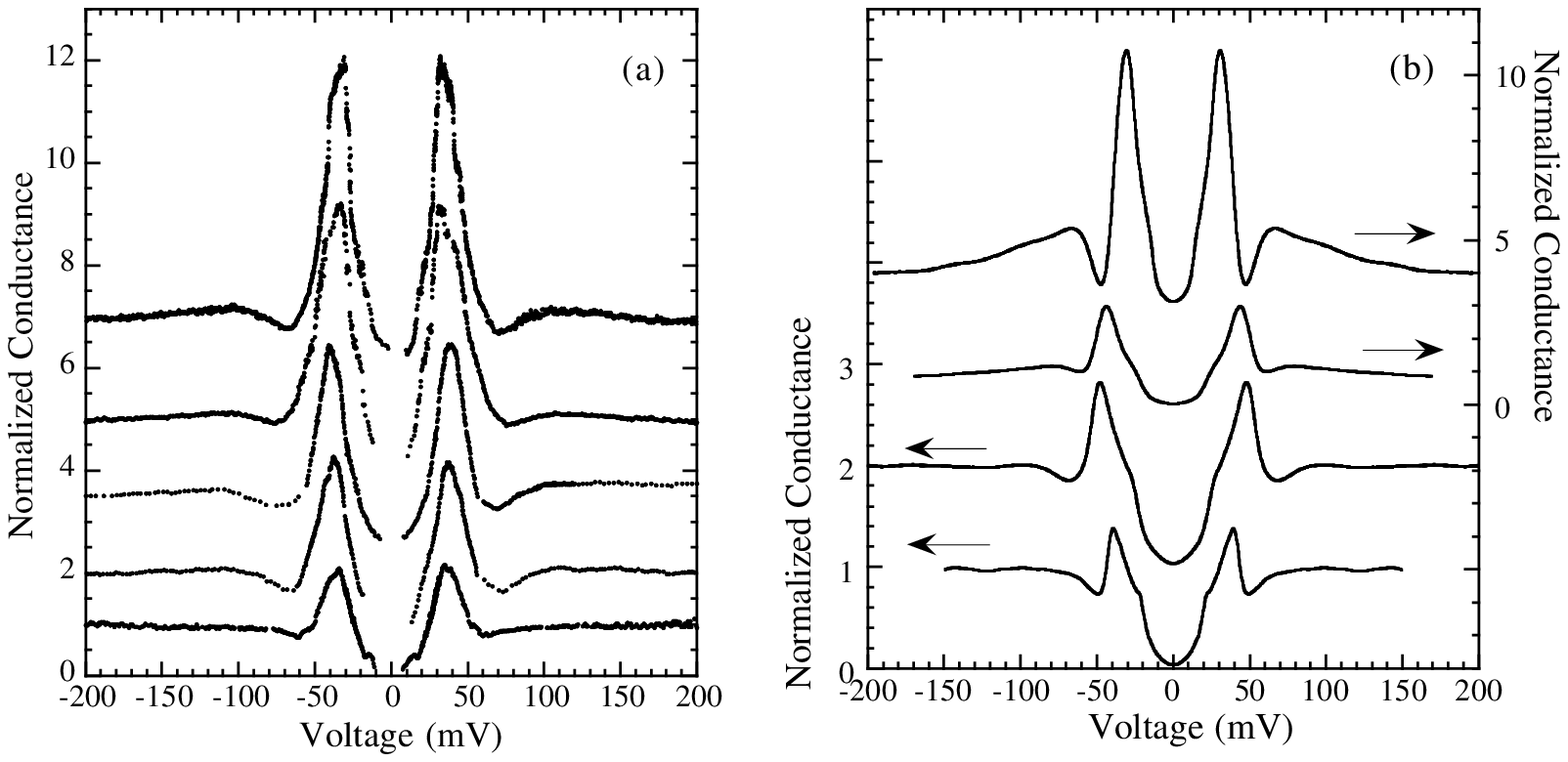}}
\vspace{-11.8cm}
\fcaption{(a) Tunneling spectra of SIS break junctions in overdoped 
B-2212 with T$_{c}$=62 K.  (b) Generated SIS tunneling spectra of 
optimally-doped Tl-2201 with T$_{c}$=86 K.}
\vspace{-0.3cm}
\end{figure} 

\end{document}